\documentclass[10pt,conference]{IEEEtran}

\usepackage{cite}
\usepackage{amsmath,amssymb,amsfonts}
\usepackage{graphicx}
\usepackage{textcomp}
\usepackage{xcolor}
\usepackage{booktabs}
\usepackage{url}

\usepackage{tikz}
\usetikzlibrary{arrows,positioning,calc,fit}
\definecolor{quanifiblue}{RGB}{226,238,247}

\usepackage{tcolorbox}
\tcbuselibrary{breakable}
\newtcolorbox{researchbox}{
   breakable,
  colback=blue!5!white,
  colframe=blue!75!black,
  fonttitle=\bfseries,
  title=Research Question,
  width=\linewidth,
  arc=2mm,
  boxrule=0.5pt
}

\providecommand{\Description}[1]{}

\begin{document}

\title{N×M-Version Programming for Quantum Software: High-Level Components across Frameworks and Engines}

\author{%
\IEEEauthorblockN{Neilson C. L. Ramalho\IEEEauthorrefmark{1},
Higor Amario de Souza\IEEEauthorrefmark{2},
Anthony Accioly\IEEEauthorrefmark{3},\\
Valter Vieira de Camargo\IEEEauthorrefmark{4},
and Marcos Lordello Chaim\IEEEauthorrefmark{1}}
\IEEEauthorblockA{\IEEEauthorrefmark{1}School of Arts, Sciences, and Humanities, University of S\~{a}o Paulo, S\~{a}o Paulo, SP, Brazil\\
neilson@usp.br, chaim@usp.br}
\IEEEauthorblockA{\IEEEauthorrefmark{2}Department of Computer and Digital Systems Engineering, Polytechnic School,\\
University of S\~{a}o Paulo, S\~{a}o Paulo, SP, Brazil\\
higoramario@usp.br}
\IEEEauthorblockA{\IEEEauthorrefmark{3}Alumni, University of S\~{a}o Paulo, S\~{a}o Paulo, SP, Brazil\\
a.accioly@alumni.usp.br}
\IEEEauthorblockA{\IEEEauthorrefmark{4}Computing Department, Federal University of S\~{a}o Carlos, S\~{a}o Carlos, SP, Brazil\\
valtervcamargo@ufscar.br}
}

\maketitle

\begin{abstract}
Quantum computing has in recent years evolved from a purely theoretical field to an active area in both academia and industry. As a result, Quantum Software Engineering has emerged as an area that aims to organize the process of building, testing, and running quantum software. However, writing quantum software still requires programming with quantum gates and qubits, as well as knowledge specific to quantum software frameworks such as Qiskit, Cirq, Qrisp, pyQuil and PennyLane, each of which reaches quantum hardware through its own toolchain. We present Quanifi, which packages the high-level routines of these frameworks as components on the Apache NiFi dataflow canvas. The components exchange circuits as OpenQASM~2.0, so a circuit built by one framework can be executed by another. This allows for two types of redundancy: several frameworks implement the same algorithm, and several engines (simulators or real hardware) execute the same circuit. We describe N$\times$M program execution, which crosses the two: a single NiFi flow runs each of the $N$ implementations on each of the $M$ engines, and because every result is labelled with its implementation and its engine, a disagreement can be linked to an implementation, to an engine, or to a single implementation-engine pair. We ran Grover's algorithm built by three frameworks on three quantum computers from IBM, IQM and Quantum Inspire, and a matrix of three adder builders on one IQM device reached through two routes. N$\times$M execution helped us discover three real defects: an incorrect gate-set declaration in the Quantum Inspire adapter, execution consistent with negated $R_x$ angles on Tuna-17, and circuit modification through Open Quantum. The latter two returned wrong answers without reporting an error. We reported all three defects and Open Quantum's provider confirmed and fixed its QASM parsing defect.
\end{abstract}


\section{Introduction}
\label{sec:intro}

Quantum Software Engineering is moving from low-level circuit construction to higher-level abstractions, in which quantum algorithms and their subroutines are treated as reusable building blocks \cite{seidel2024qrisp,leymann2019patterns}. Writing quantum software today still combines skills from different fields: quantum mechanics and linear algebra, and the APIs of frameworks such as Qiskit \cite{qiskit2024}, Cirq \cite{cirq2025}, Qrisp \cite{seidel2024qrisp} and PennyLane \cite{pennylane2018}, each of which exposes the same concepts through different method signatures, data formats and return types. This scenario is challenging for computer scientists without a physics background, similar to the challenges statistical computing created for non-programmers before packages such as SPSS \cite{wellman1998doing} appeared. This happens in both directions: classical developers might lack knowledge of quantum mechanics, while the scientists writing quantum software might often lack the engineering practices needed for creating maintainable code
\cite{DeStefano2024}.

In the classical world, low-code and visual dataflow environments helped to solve a similar problem by letting users assemble pre-built blocks on a canvas instead of writing code \cite{sahay2020lowcode, johnston2004dataflow}. 
This is how statistical analysis, data mining, instrument control, and event-driven integration became accessible to domain experts (Section~\ref{sec:rw_lowcode}). In this paper, we explore a similar approach for quantum software.

We present Quanifi, a tool that packages quantum algorithms as configurable components in the Apache NiFi dataflow platform \cite{nifi_software}, so that practitioners can assemble quantum pipelines on a visual canvas without writing circuit-level code. Quanifi preserves the concrete framework behind each component, which has the advantage of reusing parts of frameworks that are widely used and tested by their respective communities. The implementation diversity and the compatibility between components (as the exchange format for circuits is QASM2\cite{openqasm2}) let one flow run several implementations of the same routine on several simulators and quantum computers, and compare where their results differ. The contributions of this paper are:

\begin{enumerate}
  \item Quanifi, a library of 108 NiFi processors that wrap the high-level routines of Qiskit, Cirq, Qrisp, PennyLane and pyQuil, with Amazon Braket \cite{braket2026} and Microsoft Q\#/QDK \cite{qdk2026} as two further execution backends (Section~\ref{sec:quanifi}). The processors exchange circuits as OpenQASM~2.0, so a circuit built by one framework can be executed by another, and a developer can configure and connect high-level routines without implementing each algorithm from gates or learning every framework API.

  \item N$\times$M program execution, in which the $N$ framework implementations of a routine and the $M$ execution engines are combined into an $N \times M$ matrix of paths in one NiFi flow, so that a disagreement can be linked to an implementation, to an engine, or to an implementation-engine pair
  (Section~\ref{sec:nmversion}).

  \item An evaluation on quantum computers from IBM, IQM and Quantum Inspire, including two routes to the IQM device (Section~\ref{sec:hardware}). The comparisons helped us to discover three real defects: an incorrect gate-set declaration, incorrect $R_x$ execution, and circuit modification by a cloud service. We reported all three, and the cloud-service defect has been confirmed and fixed.
\end{enumerate}

This work is organised as follows. Section~\ref{sec:related} reviews low-code programming, high-level quantum languages, and multi-version execution. Section~\ref{sec:quanifi} describes Quanifi, and the NiFi component model. Section~\ref{sec:nmversion} describes N$\times$M program execution, and Section~\ref{sec:evaluation} states the research questions and the evaluation methodology. Section~\ref{sec:hardware} reports the hardware runs and the three defects. Section~\ref{sec:discussion} discusses the limitations and answers the research questions, and Section~\ref{sec:conclusion} has the conclusion of this work.

\section{Background and Related Work}
\label{sec:related}
In this section, we give an overview of the main low-code and visual programming techniques in classical computing as well as briefly describe the current developments in the quantum computing area. 
\subsection{Low-code and visual dataflow programming}
\label{sec:rw_lowcode}

Low-code development platforms provide visual, drag-and-drop interfaces in which applications are created from modular blocks that encapsulate logic, reducing the amount of hand-written code required \cite{sahay2020lowcode}. Apache NiFi follows a similar concept called dataflow programming, in which a program is a directed graph of operators that pass data along edges \cite{johnston2004dataflow}. These ideas have been used in the past to lower the technical barrier to a domain: SPSS \cite{SPSS70} and its visual Modeler made statistical analysis accessible to social scientists and business analysts who were not programmers. Similarly, KNIME \cite{knime2009} did the same for data mining and, later, machine-learning pipelines. LabVIEW \cite{kodosky2020labview} brought instrument control to scientists and engineers, and Node-RED \cite{blackstock2014nodered} allowed IoT developers to work with event-driven integration. In each case, domain experts describe their problem in their own terms by combining reusable components they did not have to write themselves.

\subsection{High-level quantum programming}
\label{sec:rw_hlq}
The need to raise the abstraction level in quantum computing has been listed as a challenge by the Quantum Software Engineering community \cite{QSERoadmapChallengesAhead, TestingDebuggingTheRoad2030}. Currently, languages and frameworks such as Silq \cite{silq2020} and Qrisp
\cite{seidel2024qrisp} provide higher-level constructs
(typed quantum variables, automatic uncomputation, algorithmic primitives) that
simplify the development of quantum programs when compared to circuit construction with gates. These approaches make it easier for the developer to create quantum programs, but
still require programming expertise. Besides, the developer needs to learn that specific language and its API. Quanifi acts then as a higher layer: it does not propose a new language, but implements and exposes the high-level primitives that these frameworks already provide as visual components (or processors, as they are called in NiFi), and unifies them behind a common interface.

\subsection{Low-code and workflow platforms for quantum computing}
\label{sec:rw_qlowcode}

There have been recent efforts to combine low-code or workflow modeling with quantum computing.
Commercial platforms such as Classiq \cite{classiq2022} offer visual or
model-driven quantum development, but often use proprietary components that are
not freely available or are bound to a single vendor. A related set of works
approaches quantum application development through patterns and workflows
\cite{leymann2019patterns}.

The work of Stiliadou et al.~\cite{stiliadou2026lowcode} is the closest to ours. It is a model-driven environment whose drag-and-drop blocks describe a quantum algorithm abstractly, and a transformer then compiles the model to OpenQASM~3 and to hybrid workflows that a cloud stack deploys across providers. Their abstraction is vendor-independent, so the model is lowered to a single canonical representation and there is, by construction, only one implementation to run. Quanifi follows a different approach: each processor wraps the routine of one concrete SDK, keeping the framework identity, and several implementations of the same concept coexist on one canvas and interoperate through OpenQASM~2.0 (Section~\ref{sec:interop}). Having the routine logic being backed by the real framework implementation also allows traceability in case there is a need to debug possible issues. As for multiple implementations of the same routine, the N$\times$M execution of Section~\ref{sec:nmversion} takes advantage of it, as components from different frameworks can be treated as independently developed versions of the same program.

\subsection{Multi-version execution}
\label{sec:rw_testing}

The output of a quantum program is a distribution over measurement outcomes. The expected distribution is often unknown, and a wrong run returns a distribution just like a correct one, which makes it difficult to decide whether a result is right \cite{TestingDebuggingTheRoad2030}. Different approaches deal with this by comparing the outputs of different versions of the same program with one another. For example, differential testing compares the output of two or more independent implementations of the same specification \cite{mckeeman1998differential}. A disagreement between the implementations might indicate that at least one of them may contain a fault, even when the expected output is not known. N-version programming follows a similar idea as it runs independently developed implementations and uses their agreement to detect faults \cite{avizienis1985nversion}. Both techniques depend on the availability of implementations that provide equivalent behavior.

N-version programming varies only the implementation. Quantum software has a second axis, because a circuit and the engine that executes it are separate artifacts: one algorithm can be built by several frameworks, and one circuit can be executed by several independently implemented simulators or by several quantum computers. Comparing each axis individually might show possible version disagreements. However, it is also possible to compare versions generated by crossing both dimensions, resulting in the matrix we describe as N$\times$M program execution in Section~\ref{sec:nmversion} and showing as well the combination that caused the disagreement.

\section{Quanifi: composing quantum circuits with Apache NiFi}
\label{sec:quanifi}

We developed Quanifi, a tool that integrates five quantum computing frameworks (IBM Qiskit \cite{qiskit2024}, Google Cirq \cite{cirq2025}, Qrisp
\cite{seidel2024qrisp}, Rigetti pyQuil~\cite{smith2017practicalquantuminstructionset}, and PennyLane \cite{pennylane2018}) with Apache NiFi
\cite{nifi_software}, an open-source dataflow orchestration platform. It additionally supports two execution-only backends, an Amazon Braket \cite{braket2026} simulator and device gateway, and a Microsoft Q\#/QDK \cite{qdk2026} simulator. Quanifi is also our proof of concept for N$\times$M program execution (Section~\ref{sec:nmversion}).

Apache NiFi builds data pipelines on a drag-and-drop canvas, where each step is a self-contained processor that consumes and produces a unit of data called a \textit{FlowFile}. Its native processors cover data-engineering work: ingestion (\textit{GetFile}, \textit{ConsumeKafka}), routing (\textit{RouteOnAttribute}), transformation (\textit{ConvertRecord}), and delivery (\textit{PutFile}, \textit{PutDatabaseRecord}). A three-step pipeline that reads CSV files from a folder, converts them to JSON, and writes the result elsewhere is assembled by placing three processors and filling in their property panels, with no code. A processor that fails routes its FlowFile to a separate failure relationship rather than losing it.
Each processor is a filter, and each queue between two processors is a pipe, which is the pipes-and-filters architectural style used in enterprise integration~\cite{Hohpe2003Enterprise}. A step (processor) performs one transformation, does not have context about its neighbours, and can be replaced by any other step that consumes and produces the same message. In Quanifi, that property is what later allows one framework's component, or one execution engine, to replace another.

In Quanifi, the FlowFile content carries the quantum artifact itself: a circuit serialised in the producing framework's format, or, after simulation, the measurement counts as JSON. The attributes carry a small metadata contract shared by all processors. A circuit builder sets \texttt{circuit.format} (\texttt{qasm2}, \texttt{qasm3}, \texttt{cirq\_json}, or \texttt{qpy}) together with circuit metrics such as \texttt{circuit.num\_qubits} and \texttt{circuit.depth}. These attributes are read by the simulator, which executes the
circuit, and returns \texttt{sim.shots}, \texttt{sim.top\_result}, and
\texttt{sim.top\_probability}. We use OpenQASM~2.0 as a bridge format
(Section~\ref{sec:interop}), so a circuit built by a Cirq processor can be executed by a Qiskit or Qrisp simulator. This attribute contract is what makes processors from different quantum frameworks composable on the same canvas.

\begin{figure*}[!htbp]
  \centering
  \includegraphics[width=\linewidth]{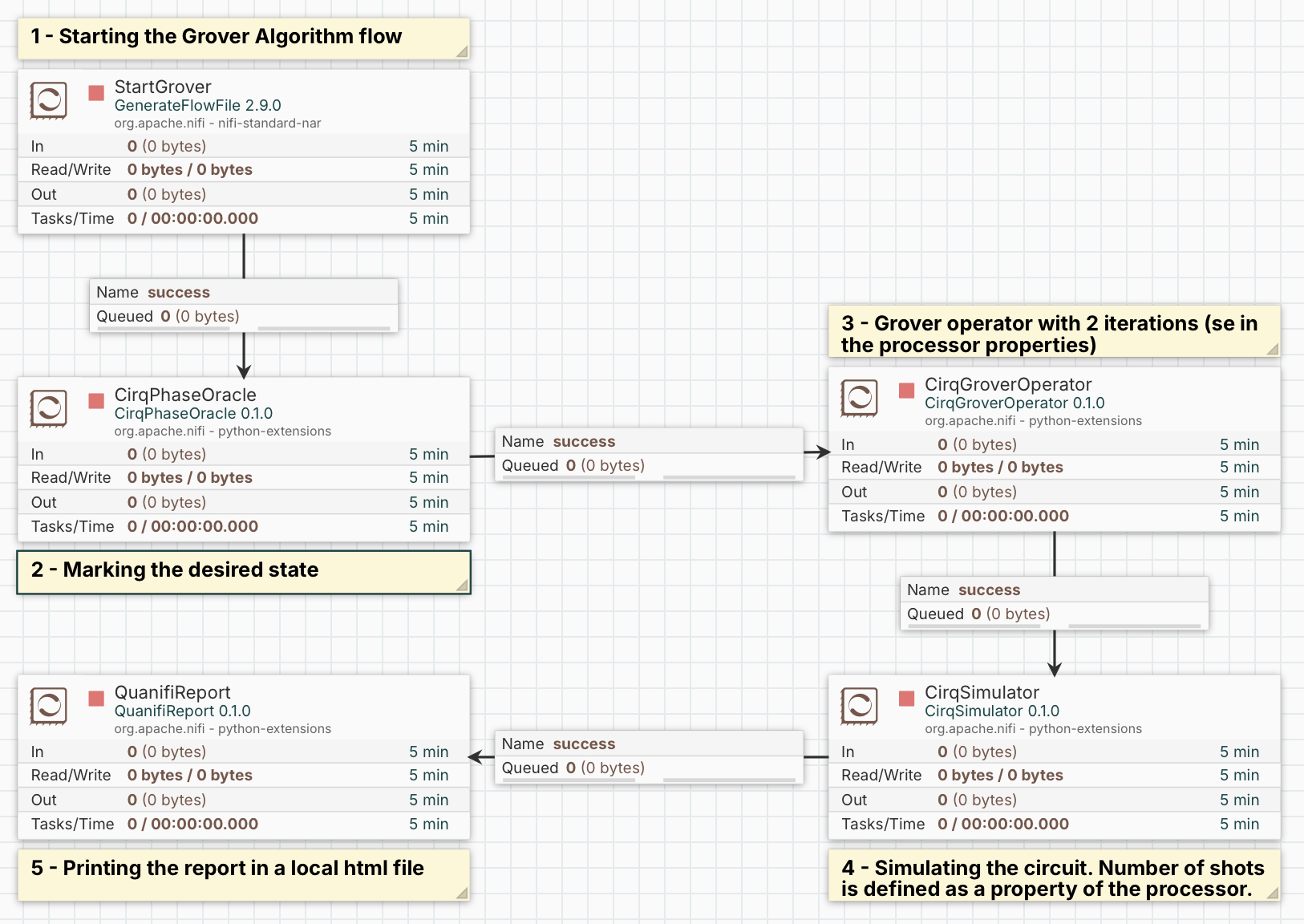}
  \caption{The Grover pipeline on the NiFi canvas, numbered in execution order.\label{fig:grover_flow}}
\end{figure*}

Quanifi adds a library of 108 processors to NiFi: 80 framework-specific processors, each encapsulating a quantum component, sub-routine, simulator, or hardware gateway from one of the frameworks, and 28 framework-agnostic processors for analysis, comparison, reporting, problem encoding, hardware submission, and testing. To run a Grover search, the user assembles a pipeline of five processors on the canvas: a standard \texttt{GenerateFlowFile} processor emits the initial FlowFile that triggers the pipeline, \texttt{CirqPhaseOracle} builds the phase oracle marking the target bitstring, \texttt{CirqGroverOperator} prepares the uniform superposition and applies the Grover operator (oracle plus diffuser) for the chosen number of iterations, \texttt{CirqSimulator} measures the resulting state, and \texttt{QuanifiReport} produces the HTML report. Each processor is configured through the NiFi UI by setting properties such as the target bitstring or the number of iterations, so no circuit-level code is written.
Figure~\ref{fig:grover_flow} shows the resulting pipeline on the NiFi canvas with the five steps. Each step passes the FlowFile with the serialised circuit or measurement counts to the next one once the simulator has run.

In terms of project structure and maintainability, Quanifi is covered by a test suite that exercises each processor in isolation (unit) and as part of end-to-end pipelines (integration). These tests run without a live NiFi instance, as we use lightweight stubs for the NiFi Java API. The tests also cover functional aspects of each algorithm. For example, the two-qubit Grover test checks that after one iteration the marked state
$|11\rangle$ is measured with probability above 95\%. 

The reusable Quanifi implementation is maintained in the framework repository~\cite{quanifiRepository}. The experiment scripts, archived results, processor snapshot, tests, and figures are collected in a separate replication package~\cite{quanifiReplication}. Its \texttt{supplementary/index.html} gallery presents additional executed algorithm flows, with canvas screenshots, processor settings, saved outputs, and independent checks for inspection without a running NiFi instance.

\subsection{The processor library}
\label{sec:library}

The 80 framework-specific processors fall into three groups: (i) circuit builders, which construct the circuit for one concept; (ii) all-in-one algorithms, which build, run, and post-process a complete routine such as VQE or QAOA; and (iii) execution processors, which run a circuit on a simulator or a quantum computer. Table~\ref{tab:capabilities} shows which frameworks have a processor for each capability. Most processors wrap a routine the framework already provides, such as Qiskit's \texttt{grover\_operator} or PennyLane's \texttt{GroverOperator}. An empty cell means that the framework offers no such routine or that Quanifi does not yet wrap it. Where a framework lacks a routine, we built the processor from its gates. That is the case for every pyQuil builder and algorithm, every Cirq builder and algorithm except the QFT and the problem Hamiltonian, and Qrisp's Deutsch--Jozsa, Bernstein--Vazirani, SWAP test and ansatz. Where a capability has several implementations, a processor from one framework can replace its equivalent from another, and N$\times$M execution (Section~\ref{sec:nmversion}) can run them side by side.

\begin{table}[!htbp]
\centering
\caption{Capabilities of the 80 framework-specific processors. A check mark means a processor exists for that capability and framework.}
\label{tab:capabilities}
{\footnotesize\setlength{\tabcolsep}{2.5pt}%
\begin{tabular}{@{}lccccc@{}}
\toprule
Capability & Qiskit & Cirq & Qrisp & PennyLane & pyQuil \\
\midrule
\multicolumn{6}{@{}l}{\textbf{Circuit builders}}\\
Hadamard transform          & \checkmark & \checkmark & \checkmark &   &   \\
State preparation           & \checkmark & \checkmark & \checkmark &   &   \\
Typed data encoding         &   &   & \checkmark &   &   \\
Phase oracle                & \checkmark & \checkmark &   &   & \checkmark \\
Grover operator             & \checkmark & \checkmark &   &   & \checkmark \\
Grover circuit              & \checkmark & \checkmark &   & \checkmark & \checkmark \\
Amplitude amplification     & \checkmark &   &   &   &   \\
Quantum Fourier transform   & \checkmark & \checkmark & \checkmark &   & \checkmark \\
Phase estimation            & \checkmark & \checkmark & \checkmark &   &   \\
Truth-table synthesis       &   &   & \checkmark &   &   \\
Problem Hamiltonian         & \checkmark & \checkmark & \checkmark &   &   \\
Variational ansatz          & \checkmark & \checkmark & \checkmark & \checkmark & \checkmark \\
QAOA circuit (fixed angles) &   &   &   &   & \checkmark \\
Quantum arithmetic (adders) & \checkmark & \checkmark & \checkmark & \checkmark &   \\
Feature embedding (QML)     &   &   &   & \checkmark &   \\
Dataset loader (QML)        &   &   &   & \checkmark &   \\
Bell, teleportation, GHZ    & \checkmark &   &   &   &   \\
\midrule
\multicolumn{6}{@{}l}{\textbf{All-in-one algorithms}}\\
Grover search               & \checkmark &   & \checkmark &   &   \\
VQE                         & \checkmark & \checkmark & \checkmark &   & \checkmark \\
QAOA                        & \checkmark & \checkmark & \checkmark & \checkmark & \checkmark \\
Amplitude estimation        & \checkmark & \checkmark & \checkmark & \checkmark &   \\
Deutsch--Jozsa              &   &   & \checkmark &   &   \\
Bernstein--Vazirani         &   &   & \checkmark &   &   \\
SWAP test                   &   &   & \checkmark &   &   \\
Shor factoring              &   &   & \checkmark &   &   \\
Variational classifier (QML) &  &   &   & \checkmark &   \\
\midrule
\multicolumn{6}{@{}l}{\textbf{Execution}}\\
Shot-based simulator        & \checkmark & \checkmark & \checkmark & \checkmark & \checkmark \\
Statevector simulator       & \checkmark & \checkmark &   &   &   \\
Expectation values          &   &   &   & \checkmark & \checkmark \\
Quantum computer            & \checkmark &   & \checkmark &   &   \\
\bottomrule
\end{tabular}}
\par\smallskip\raggedright\footnotesize
Hardware: IBM via Qiskit and IQM via Qrisp, plus batched submitters for IBM, IQM and Quantum Inspire. Amazon Braket and a Q\#/QDK simulator provide further engines.
\end{table}

\subsection{Cross-framework interoperability via OpenQASM 2.0}
\label{sec:interop}

Quanifi has processors to build circuits from five different frameworks, plus processors to integrate with backends for execution from Amazon Braket and Microsoft Q\#/QDK. The frameworks vary in which circuit formats they support. For example,  QASM3 \cite{openqasm3} and QPY are used by Qiskit, and \texttt{cirq\_json} is specific to Cirq. 
Although OpenQASM~3 is widely adopted as an export target, support for importing it across frameworks is still partial~\cite{openqasmimplementations}. Qiskit describes its importer as being ``in its infancy'' and its native parser as experimental, with a reduced feature set~\cite{qiskitqasm3docs}. In an independent benchmark, that native parser accepted 2 of 11 OpenQASM~3 test programs, against 11 of 11 for the reference parser~\cite{kim2024qasmts}. Cirq's importer is also experimental and supports only a subset of the specification~\cite{cirqinteropdocs}.  We use OpenQASM~2.0 (\texttt{QASM2})~\cite{openqasm2} as the interchange format, because every circuit builder in Quanifi can emit it and every simulator and hardware gateway accepts it. Setting a circuit processor's output format to \texttt{QASM2} therefore allows its output to be processed by any of the simulators. For example, a Cirq-built Grover circuit can run on the Qiskit Aer simulator, and a Qiskit QFT circuit can be run on an IQM backend. The Amazon Braket backend consumes the same circuits (natively OpenQASM~3, with \texttt{QASM2} imported on input), and the Q\#/QDK backend accepts OpenQASM~2.0 and~3 directly.

Amazon Braket and Microsoft Q\#/QDK appear in Quanifi only as execution engines. Both accept circuits from the other frameworks, but neither is a practical source of circuits for them. Q\# compiles to QIR (Quantum Intermediate Representation) and cannot export OpenQASM. Braket has an algorithm library~\cite{braketAlgorithmLibrary}, but its circuits are Braket objects that export only to Braket's own OpenQASM~3 dialect, with gate names such as \texttt{cnot} and the phase behaviour described below.

Passing a circuit between frameworks needed two adjustments, both caused by the way QASM separates a gate's name from its definition. A QASM program body is a list of gate applications written as \textit{names}, but only a small set of primitives is built into the language, and the other standard gates are defined in terms of those primitives in an include file. OpenQASM~2.0 builds in \texttt{U} and \texttt{CX} and puts the rest in \texttt{qelib1.inc}. OpenQASM~3.0 builds in \texttt{U} and \texttt{gphase} and uses \texttt{stdgates.inc}. 
There are cases in which the receiving framework does not support the gates or definitions generated by the producing framework. For example,  Qiskit's default OpenQASM~2 parser does not recognise all the additional gate names emitted by other frameworks. Our processors enable \texttt{qasm2.LEGACY\_CUSTOM\_INSTRUCTIONS}, Qiskit's compatibility mapping for these names.  In another case, Amazon Braket reads OpenQASM~3 but does not include a \texttt{stdgates.inc}, and pasting a copy of those definitions into the program does not work: they are written in terms of \texttt{U} and \texttt{gphase}, and the phase Braket's interpreter applies for these primitives did not match the one described in the OpenQASM~3 specification~\cite{openqasm3}. A global phase is unobservable on an isolated gate, but under a control it becomes a relative phase between the two branches, which is the phase-kickback mechanism that Grover's diffuser and phase estimation use. These definitions, which are correct in isolation, fail once they are used in controlled gates: circuits with inlined definitions are sampled correctly on their own, but a four-qubit Grover circuit splits its peak between the marked state and its neighbour. Our Amazon Braket processors avoid gate definitions altogether: they transpile each incoming circuit down to the small set of gates Amazon Braket implements natively and emit only those names. For the circuits we tested, we avoided these issues by emitting only standard gate names that each target implements, as the imported definitions depend on how the target interprets their primitives.

Using \textit{QASM2} as the bridge format also allows for the N$\times$M execution of Section~\ref{sec:nmversion}, as it lets one implementation replace another, and every engine, simulator, or quantum computer receive the circuit of every implementation.

\section{N$\times$M program execution}
\label{sec:nmversion}
\begin{figure}[!t]
  \centering
  \begin{tikzpicture}[
    font=\scriptsize,
    blk/.style={draw=black!60, fill=quanifiblue, rounded corners=1.5pt,
                align=center, inner sep=2.5pt, text width=2.2cm, minimum height=8mm},
    wide/.style={blk, text width=7.6cm},
    note/.style={font=\scriptsize\itshape, text=black!65, align=left, inner sep=1pt},
    ar/.style={->, >=stealth, draw=black!55, line width=0.4pt}]

    \def\cA{1.3} \def\cB{4.0} \def\cC{6.7}

    \node[wide] (src) at (\cB,0)
      {\textbf{QuantumTestCaseSource} $\rightarrow$ \texttt{SplitJson}\\
       four cases: marked states \texttt{10}, \texttt{11}, \texttt{0111}, \texttt{0110}};

    \node[blk] (b1) at (\cA,-1.55) {\textbf{Qiskit}\\Grover};
    \node[blk] (b2) at (\cB,-1.55) {\textbf{PennyLane}\\Grover};
    \node[blk] (b3) at (\cC,-1.55) {\textbf{Cirq}\\Grover};

    \node[blk] (d1) at (\cA,-3.4) {\textbf{IBM}\\\texttt{ibm\_kingston}\\submit $\rightarrow$ poll};
    \node[blk] (d2) at (\cB,-3.4) {\textbf{IQM}\\\texttt{garnet}\\submit $\rightarrow$ poll};
    \node[blk] (d3) at (\cC,-3.4) {\textbf{Quantum Inspire}\\\texttt{Tuna-17}\\submit $\rightarrow$ poll};

    \node[wide] (or) at (\cB,-4.95)
      {\textbf{QuantumSuccessProbabilityOracle}\\
       compares the nine builder--device results of each case};
    \node[wide, minimum height=5.5mm] (rep) at (\cB,-6.0) {\textbf{QuanifiReport}};

    \foreach \b in {b1,b2,b3}{
      \draw[ar] (src) -- (\b);
      \foreach \d in {d1,d2,d3}{\draw[ar] (\b.south) -- (\d.north);}
    }
    \foreach \d in {d1,d2,d3}{\draw[ar] (\d) -- (or);}
    \draw[ar] (or) -- (rep);

  \end{tikzpicture}
  \caption{The version matrix as it runs on quantum computers: one case, three Grover builders, three devices, and the nine resulting paths meeting at one oracle.}
  \Description{A flow diagram: a test case source feeds three Grover builder processors (Qiskit, PennyLane, Cirq); each builder connects directly to all three device gateways (IBM ibm_kingston, IQM garnet, Quantum Inspire Tuna-17), each with a submit and a poll processor, giving nine builder-device paths; all nine paths converge on a success probability oracle and then a report processor.}
  \label{fig:version_matrix_flow}
\end{figure}
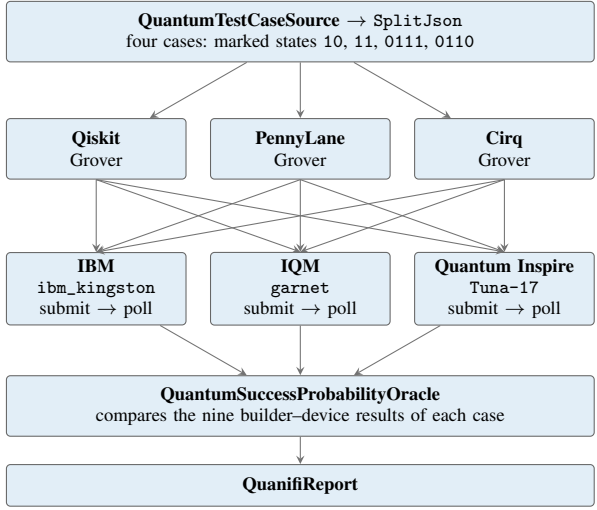

NiFi can copy a processor's output to several downstream processors. Quanifi uses this to run circuits from each of $N$ builders on each of $M$ execution engines, giving $N \times M$ paths per input case. The builders exchange circuits as OpenQASM~2.0, and the engines can be simulators or quantum computers.
We call the grid of paths the \textit{version matrix}: rows are builders, and columns are engines. Figure~\ref{fig:version_matrix_flow} shows the version matrix of our first hardware experiment (Section~\ref{sec:evaluation}): circuits from three Grover builders (Qiskit, PennyLane and Cirq) run on three devices, giving nine paths per case.

\begin{figure}[!htbp]
  \centering
  \includegraphics[width=\linewidth]{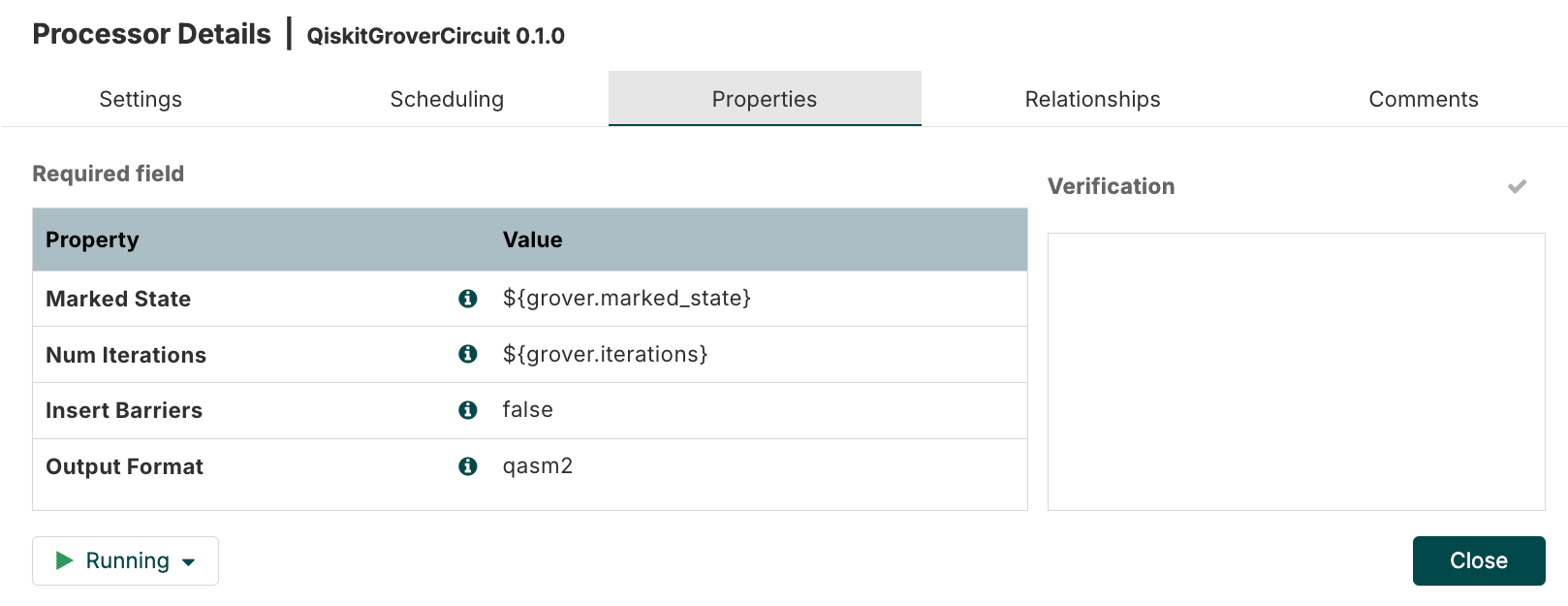}
  \caption{Properties of the \texttt{QiskitGroverCircuit} processor. The marked state and iteration count are Expression Language references, so each input case sets them through its FlowFile attributes.}
  \Description{Screenshot of the NiFi Processor Details dialog for QiskitGroverCircuit, with Marked State set to grover.marked_state and Num Iterations set to grover.iterations as Expression Language references, Insert Barriers false, and Output Format qasm2.}
  \label{fig:expression_language}
\end{figure}

A \texttt{GenerateFlowFile} processor starts the flow. \texttt{QuantumTestCaseSource} supplies a table of input cases, and \texttt{SplitJson} creates one FlowFile per case. Processor properties, such as the marked state and iteration count, read their values from that FlowFile's attributes using NiFi Expression Language (for example, \texttt{\$\{grover.marked\_state\}}, Figure~\ref{fig:expression_language}). Each case goes to all three builders. Qiskit's transpiler then maps each builder's circuit to each device's gates and qubits. The nine results meet at \texttt{QuantumSuccessProbabilityOracle}, which compares the probability of measuring the marked state.

Each result retains its builder and engine as FlowFile attributes. Failures across a row point to one builder, failures across a column to one engine, and a failure in one cell to a particular builder--engine pair. This comparison helps discover bugs by checking both whether a builder's circuit works on other engines and whether other builders work on the same engine. Varying only one axis provides only one of these checks. Targeted tests can then investigate the cause. A case requires $N \times M$ executions instead of $N$ or $M$. Comparing every pair of results would require $NM(NM-1)/2$ comparisons, or 36 for the nine paths in Figure~\ref{fig:version_matrix_flow}.

Hardware jobs may wait in a provider's queue for hours. Each device therefore has a submission processor and a polling processor, with the job identifier carried in the FlowFile. NiFi queues retain pending jobs, back-pressure limits submissions, and failed jobs follow a separate failure relationship. NiFi's provenance records the builder, device, and batch for each result.

Section~\ref{sec:evaluation} describes how we evaluated N$\times$M execution on quantum computers, and Section~\ref{sec:hardware} reports the results.

\section{Evaluation methodology and research questions}
\label{sec:evaluation}

The evaluation is organised around two research questions:

\noindent\textbf{RQ1 (Component-level expressiveness).} What quantum programs can be expressed at the component level, without writing framework-specific circuit code?

\noindent \textbf{RQ2 (Attribution and cost of the version matrix).} What does running every combination of framework implementation and execution engine on real quantum computers reveal that comparing implementations alone, or engines alone, does not, and what does it cost?

\textbf{Experimental setup.} We address RQ1 by building flows from existing processors and 
checking that no framework-specific circuit code is needed: the Grover flows of the hardware experiments below,
 and the supplementary workflows in the replication package~\cite{quanifiReplication}. We address RQ2 with two experiments on real quantum computers.

\textit{Experiment~1: three builders on three devices.} We ran the version matrix of Figure~\ref{fig:version_matrix_flow} on IBM's \texttt{ibm\_kingston} (156 qubits), IQM's \texttt{garnet} (20 qubits) and Quantum Inspire's \texttt{Tuna-17} (17 qubits). Three builders (Qiskit, PennyLane and Cirq) produced Grover circuits for the two-qubit marked states \texttt{10} and \texttt{11} and the four-qubit states \texttt{0111} and \texttt{0110}, and each circuit ran for 1024 shots. Each batch also contains two controls per case. A readout baseline prepares the marked state directly and measures it, showing how often the device returns that state without the search circuit. A null replicate is an exact copy of the Qiskit circuit, showing how much two identical runs differ. A batch therefore has 20 circuits per device: 12 builder circuits, four readout baselines, and four null replicates. We repeated the batches on each device, both on the same calibration and on later ones.

\textit{Experiment~2: two routes to one device.} We reached \texttt{garnet} both directly, through IQM Resonance, and through the Open Quantum cloud service~\cite{openquantum2026}, so that the engine axis has two routes to the same device. Both routes received identical circuit files at the same time, with 512 shots per circuit: three one-bit adders (Qiskit's CDKM ripple-carry, Cirq's QFT adder, and PennyLane's out-of-place adder) on four inputs, a single Toffoli gate, and the Grover circuits and readout baselines of Experiment~1. This experiment followed an earlier adder run on Rigetti's \texttt{Cepheus-1-108Q} through Open Quantum, described in Section~\ref{sec:hardware}.

\textbf{Metrics.} For Grover, the success probability is the number of shots that return the marked state divided by the total number of shots. A random guess succeeds with probability $0.25$ in the two-qubit cases and $0.0625$ in the four-qubit cases. For the adders and the Toffoli gate, we report the probability of the correct answer and, when the answer is wrong, the most frequent outcome.

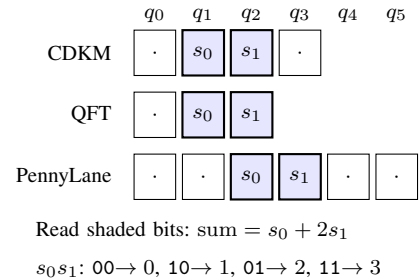
\begin{figure}[!htbp]
\centering
{
\begin{tikzpicture}[
  font=\footnotesize,
  bit/.style={draw=black, minimum width=5.5mm, minimum height=6mm, inner sep=0pt},
  sum bit/.style={bit, fill=blue!10, line width=0.8pt}]
  \foreach \i in {0,...,5}{
    \node at (0.64*\i,0.50) {$q_{\i}$};
  }
  \node[anchor=east] at (-0.45,0) {CDKM};
  \node[anchor=east] at (-0.45,-0.82) {QFT};
  \node[anchor=east] at (-0.45,-1.64) {PennyLane};
  \foreach \i in {0,3}{\node[bit] at (0.64*\i,0) {$\cdot$};}
  \node[sum bit] at (0.64,0) {$s_0$};
  \node[sum bit] at (1.28,0) {$s_1$};
  \node[bit] at (0,-0.82) {$\cdot$};
  \node[sum bit] at (0.64,-0.82) {$s_0$};
  \node[sum bit] at (1.28,-0.82) {$s_1$};
  \foreach \i in {0,1,4,5}{\node[bit] at (0.64*\i,-1.64) {$\cdot$};}
  \node[sum bit] at (1.28,-1.64) {$s_0$};
  \node[sum bit] at (1.92,-1.64) {$s_1$};
  \node[anchor=west] at (-1.7,-2.35) {Read shaded bits: $\mathrm{sum}=s_0+2s_1$};
  \node[anchor=west] at (-1.7,-2.85) {$s_0s_1$: \texttt{00}$\to 0$, \texttt{10}$\to 1$, \texttt{01}$\to 2$, \texttt{11}$\to 3$};
\end{tikzpicture}}
\caption{Sum-bit positions in the one-bit adder experiments, with $q_0$ on the left. Shaded boxes contain the sum. Dots denote other measured bits, which are ignored when scoring the sum. Each row shows the full measured register.}
\Description{Three rows of measured qubit registers, one per adder (CDKM with four qubits, QFT with three, PennyLane with six), with the two sum bits shaded: q1 and q2 for CDKM and QFT, q2 and q3 for PennyLane. The sum is s0 plus two times s1.}
\label{fig:adder_output_bits}
\end{figure}

For the one-bit adders on \texttt{garnet}, we read the sum from two positions in the measured bit string, shown in Figure~\ref{fig:adder_output_bits}. We counted a shot as successful if those two bits encoded the expected sum, whatever values appeared in the other positions. For example, the CDKM outcomes \texttt{0100} and \texttt{1100} both have \texttt{10} in the sum positions and therefore both count towards a sum of 1. The success probability is the number of such shots divided by the total number of shots. The first sum bit is the least significant: \texttt{10} represents 1, and \texttt{01} represents 2. There are four possible two-bit outcomes, so a uniform random guess gives the expected sum with probability $1/4$. For the standalone Toffoli check, all three measured bits must match the expected outcome \texttt{111}.

\textbf{Gate-count analysis.} Two-qubit gates are a main source of error on current devices, so a builder whose routed circuit needs fewer two-qubit gates may reach a higher success probability. If it does, differences between builders on a device reflect circuit cost rather than a defect. We tested this for the four-qubit cases, where the routed circuits of different builders differ by up to 58 two-qubit gates on the same device (Table~\ref{tab:hardware_matrix}). On each device, we ordered the builders from fewest to most two-qubit gates, and compared the first with the second and the second with the third, each with a one-sided two-proportion $z$-test. We applied a Holm correction at a significance level of $0.05$ and also required an observed difference of at least $0.10$. A case supports the ordering only if both comparisons meet these conditions. The two-qubit circuits are small, and we report their success probabilities without a test.

\textbf{Cost.} For RQ2, we measure the cost of N$\times$M execution as the number of circuit runs and shots it requires, including the controls. We did not measure financial cost or waiting time.

Section~\ref{sec:hardware} reports the results of both experiments.

\section{Execution on quantum computers}
\label{sec:hardware}

\textbf{Three builders on three quantum computers.} We ran Experiment 1 (Section \ref{sec:evaluation}) on three quantum computers: IBM's \texttt{ibm\_kingston} (156 qubits), IQM's \texttt{garnet} (20 qubits) and Quantum Inspire's \texttt{Tuna-17} (17 qubits). Three builders (Qiskit, PennyLane and Cirq) produced circuits for the two-qubit marked states \texttt{10} and \texttt{11} and the four-qubit states \texttt{0111} and \texttt{0110}, and each circuit ran for 1024 shots. On hardware, a circuit must first be routed onto the device's qubits. All three builders share Qiskit's transpiler for this step, so the transpiler is the same for every builder: the builders' raw circuits already differ (56, 88, and 112 two-qubit gates for the four-qubit cases, for Qiskit, PennyLane and Cirq), optimisation leaves these counts unchanged, and routing adds between 18 and 96 gates depending on builder and device, enough to
reorder the builders by gate count on \texttt{garnet} and \texttt{Tuna-17}. Each batch also contains a readout baseline, which prepares the marked state directly and measures it, and a null replicate, an exact copy of the Qiskit circuit that shows how much two identical runs differ. A batch therefore has 20 circuits per device: 12 builder circuits, four readout baselines, and four null replicates.

Table~\ref{tab:hardware_matrix} reports the first valid batch on each device (for \texttt{Tuna-17}, the first batch after the gate-set correction described below), with each builder's routed two-qubit gate count for the four-qubit cases in the last row for each device. The two-qubit circuits worked well everywhere: every builder reached a success probability between 0.92 and 0.96, within 0.05 of the readout baseline. We compared the builders using the gate-count analysis of Section \ref{sec:evaluation}. 

For the cases with four qubits, the fact that the generated circuit had fewer two-qubit gates did not consistently give a higher success probability. On \texttt{ibm\_kingston}, Qiskit, the builder with the fewest gates, had better results than PennyLane (0.16 difference) on \texttt{0110}, the only difference above the 0.10 margin. However, Cirq had the most gates and performed better than PennyLane in both cases. On \texttt{garnet} and \texttt{Tuna-17}, the four-qubit circuits were close to random guessing (0.05 to 0.12, against 0.0625), so it was not possible to compare the builders. 

Repeating the batches showed that the \texttt{ibm\_kingston} result did not hold. Its first three repeats, which ran together on a single calibration, still put Qiskit ahead of PennyLane on \texttt{0110} (by 0.08 to 0.12). But three later batches, each on a new calibration and a different set of qubits, removed the gap: no two builders differed by more than 0.04, and each of the three was ahead at least once. On \texttt{garnet}, the repeats on the same calibration and a later batch on a new one gave the same picture as the first batch, with four-qubit success between 0.08 and 0.15. PennyLane, whose circuits need the most two-qubit gates on this device, was lowest in every \texttt{garnet} batch, but never by 0.10. On \texttt{Tuna-17}, three repeats taken up to six and a half hours later matched the first batch, with two-qubit success between 0.91 and 0.95 and four-qubit success between 0.05 and 0.09.

In the batches shown in Table~\ref{tab:hardware_matrix}, the absolute differences in success probability between each Qiskit circuit and its identical null replicate ranged from 0.005 to 0.019 on \texttt{ibm\_kingston}, 0.004 to 0.012 on \texttt{garnet}, and 0.004 to 0.018 on \texttt{Tuna-17}. Thus, all twelve differences were below 0.02. Each circuit had only one duplicate in its batch, so these results describe variation within those batches and do not establish stability across calibrations.

\textbf{First defect: a device that applies a rotation in the wrong direction.} The matrix also found a real bug. In the first \texttt{Tuna-17} batch, Cirq’s circuits returned incorrect answers for both two-qubit cases. For the marked state \texttt{11}, the device returned \texttt{00} in 94\% of the shots, and for \texttt{10} it returned \texttt{01}. Every bit of the answer had been flipped, and the success probability was 0.006 and 0.012. In the same batch, Qiskit and PennyLane returned 0.93 to 0.95 on that device, and Cirq reached 0.93 to 0.96 on the other two devices. Every result has the name of its builder and its device, so the failure belongs to one cell of the matrix: Cirq on \texttt{Tuna-17}.

We simulated the circuit and checked its results for each stage, from the output of the Cirq processor to the cQASM \cite{khammassi2018cqasmv10commonquantum,cqasm3spec} program sent to the device. The results were always the correct marked state. We found out that the difference was in the gates that reached the device.
In its QASM output, Cirq uses the gate \texttt{u3} (the general single-qubit rotation) with a different set of angles for each operation such as $H$ or $X$, while Qiskit and PennyLane write the named gates directly. Before a circuit runs, it is translated into the gates that the device declares, and for \texttt{Tuna-17} the Quantum Inspire adapter declares
$R_x$ among them. The named gates of Qiskit and PennyLane were already on that list and passed through as they were, while every \texttt{u3} of Cirq had to be rewritten, and the translation used $R_x$. Cirq's circuit was therefore the only one that arrived at the device with $R_x$ gates in it. It was also the longest, with 17 gates against 12 for Qiskit and 11 for PennyLane, all three with three two-qubit gates.

Two further runs showed that \texttt{Tuna-17} applies $R_x$ with the opposite angle. Negating the $R_x$ angles in Cirq's circuit brought the marked state back with probability 0.93, and a three-gate circuit whose outcome depends on the sign of $R_x$ gave the opposite of its simulation in 93.5\% of the shots. We reported the rotation behaviour with reproducible probes~\cite{qiRxIssue395}. The matrix helped uncover it by showing that Cirq's circuits worked on other devices and that other builders worked on \texttt{Tuna-17}.

\textbf{Second defect: an adapter that advertises unsupported gates.} During that investigation, we found that the Quantum Inspire adapter builds its target from a static gate mapping instead of the device's declared gate set. It therefore advertises \texttt{cx}, \texttt{cp}, and \texttt{swap} for \texttt{Tuna-17}, whose declared two-qubit gate set contains only \texttt{CZ}. We reported this separate adapter defect~\cite{qiGateSetIssue394}. Both Quantum Inspire reports are open, without maintainer confirmation, as of 20~September~2026. To work around both defects, we transpiled the circuits using the device's declared gate set, excluding $R_x$ because of the rotation defect. Cirq's two-qubit success probability then reached 0.93, while those of the other builders changed by at most 0.02.

\begin{table}[!htbp]
\centering
\caption{Grover success probability on three quantum computers, first valid batch per device, 1024 shots per circuit.}
\label{tab:hardware_matrix}
{\footnotesize%
\setlength{\tabcolsep}{3pt}%
\begin{tabular}{llrrrr}
\toprule
Device & Case & Readout & Qiskit & PennyLane & Cirq \\
\midrule
\texttt{ibm\_kingston} & \texttt{10} (2q)   & 0.975 & 0.958 & 0.954 & 0.958 \\
                       & \texttt{11} (2q)   & 0.983 & 0.963 & 0.952 & 0.944 \\
                       & \texttt{0111} (4q) & 0.942 & 0.523 & 0.451 & 0.481 \\
                       & \texttt{0110} (4q) & 0.955 & 0.543 & 0.384 & 0.457 \\
                       & 2q gates (4q)      & --    & 121   & 131   & 146   \\
\midrule
\texttt{garnet}        & \texttt{10} (2q)   & 0.969 & 0.929 & 0.942 & 0.926 \\
                       & \texttt{11} (2q)   & 0.956 & 0.920 & 0.930 & 0.938 \\
                       & \texttt{0111} (4q) & 0.946 & 0.097 & 0.072 & 0.099 \\
                       & \texttt{0110} (4q) & 0.958 & 0.119 & 0.073 & 0.094 \\
                       & 2q gates (4q)      & --    & 126   & 184   & 130   \\
\midrule
\texttt{Tuna-17}       & \texttt{10} (2q)   & 0.960 & 0.924 & 0.942 & 0.930 \\
                       & \texttt{11} (2q)   & 0.954 & 0.930 & 0.949 & 0.933 \\
                       & \texttt{0111} (4q) & 0.913 & 0.067 & 0.082 & 0.056 \\
                       & \texttt{0110} (4q) & 0.935 & 0.061 & 0.095 & 0.055 \\
                       & 2q gates (4q)      & --    & 122   & 137   & 130   \\
\bottomrule
\end{tabular}}
\end{table}

\textbf{Third defect: a cloud service that changes the circuit.} In our experiments, we also used Open Quantum~\cite{openquantum2026}, a cloud service that offers one interface for devices of several vendors, among them Rigetti's \texttt{Cepheus-1-108Q} and IQM's \texttt{garnet}. We used it as a second way to reach a device we already reach directly (garnet). The engine axis then has two routes to \texttt{garnet}, which lets us separate the device from the service in front of it.

We found the defect while running a ripple-carry adder~\cite{cuccaro2004newquantumripplecarryaddition} on \texttt{Cepheus-1-108Q}, compiled by Qiskit to low-level gates before submission. We simulated the circuit and it returned the right sum for every input. On hardware, we ran it on seven inputs, from $0+0$ to $3+3$. Five of them returned the right sum, each input produced a dominant outcome, and re-running three of the inputs four times each gave those same results again. The run looked like an ordinary hardware result with two wrong outputs due to noise, and we decided to investigate it further, as these results had one specific wrong state. In this case, $3+0$ gave \texttt{111010} in 44 to 48\% of the shots across four repeats, while the correct \texttt{111100} appeared in at most 3 of 512 shots. With $q_0$ on the left and the sum in $q_2$, $q_3$ and $q_4$ (least significant first), \texttt{111010} encodes 5 instead of 3, and the dominant outcome for $1+3$, \texttt{100000}, encodes 0 instead of 4. We then compiled the same adder to the gates $\{H, X, \mathrm{CX}, R_z\}$, removed every $R_z$ gate, and simulated the result on our own machine. That simulation gave the device's output for all seven inputs: the wrong sums 5 and 0 for the two failing inputs, and the correct sum for the other five. This suggested that the service had changed the circuit, with results consistent with the loss of its $R_z$ rotations. In this gate set a Toffoli gate is a sequence of CNOT and Hadamard gates with $R_z$ rotations between them. Without the rotations, these gates cancel and the target qubit is not inverted.

We then ran Experiment 2 (Section\ref{sec:evaluation}), comparing two routes to \texttt{garnet}: directly through IQM Resonance and through Open Quantum. We used four inputs, from $0+0$ to $1+1$, with three adders: Qiskit's CDKM ripple-carry, Cirq's QFT, and PennyLane's out-of-place. For each component, we compiled the adder once to the gates $\{H, X, \mathrm{CX}, R_z\}$, then added the $X$ gates that prepare each input. All four inputs therefore shared the same compiled adder body, preventing input-specific compiler optimisations from changing the comparison. For the CDKM adder, we also kept a version with its two Toffoli gates undecomposed, to test whether the failure depended on the circuit's representation. We ran each adder version on all four inputs through both routes. We also tested a single Toffoli gate, submitted both as \texttt{ccx} and as its decomposition. Both routes received identical circuit files at the same time, with 512 shots per circuit.

Table~\ref{tab:oq_defect} shows the Open Quantum results. Through IQM Resonance, every circuit returned the correct answer as its most frequent outcome, with probabilities between 0.73 and 0.88. The two CDKM versions differed by at most 0.02. Through Open Quantum, the decomposed CDKM adder gave a wrong sum for $1+0$, and PennyLane's adder returned \texttt{00} for every input except $0+0$, with probabilities of 0.99 to 1.00. The QFT adder's probability of a correct answer fell to 0.22--0.31, close to random guessing (0.25). Circuits with undecomposed Toffoli gates gave the correct answer on both routes. No job reported an error or warning. Several wrong answers even had higher probabilities than the correct answers returned through the direct route. Since both routes used the same files and device at the same time, the comparison points to a fault in the Open Quantum route when handling decomposed circuits.

A simple simulation reproduces every row of Table \ref{tab:oq_defect}: remove every $R_z$ gate from the compiled body, keep the $X$ gates that prepare the inputs, and simulate the remaining circuit. This predicts both the wrong answers and the inputs for which the decomposed CDKM adder still succeeds. However, matching these results does not establish what the service changed. Open Quantum does not return the executed circuit, and the Grover experiments below include a case that this rule predicts incorrectly. The full table is in the replication package~\cite{quanifiReplication}.

We reported the defect with the circuits and job identifiers. The provider confirmed it, traced it to a QASM parsing function, and fixed it on 17~September~2026. All Open Quantum results reported here were collected before the fix.

\begin{table}[!htbp]
\centering
\caption{Open Quantum results on \texttt{garnet}.}
\label{tab:oq_defect}
{\footnotesize\setlength{\tabcolsep}{4pt}%
\begin{tabular}{@{}lllr@{}}
\toprule
Circuit & Case & Outcome & $p$ \\
\midrule
Toffoli         & decomposed            & wrong, \texttt{110} & 0.97 \\
Toffoli         & intact                & correct             & 0.83 \\
\addlinespace
CDKM adder      & decomposed, $1+0$     & wrong, \texttt{11}  & 0.86 \\
CDKM adder      & decomposed, other three & correct           & 0.88--0.93 \\
CDKM adder      & intact, all four      & correct             & 0.70--0.92 \\
\addlinespace
PennyLane adder & decomposed, $0+0$     & correct             & 1.00 \\
PennyLane adder & decomposed, other three & wrong, \texttt{00} & $\ge$0.99 \\
QFT adder       & decomposed, all four  & correct, at chance  & 0.22--0.31 \\
\bottomrule
\end{tabular}}
\end{table}

\textbf{Grover through the two routes.} We repeated the comparison with the Grover circuits in Table~\ref{tab:hardware_matrix} and the readout baselines. We compiled them to the same gate set used for the adders and sent identical files to \texttt{garnet} through both routes at the same time, with 512 shots per circuit. Through IQM Resonance, the two-qubit circuits from all three builders returned the correct answer with probabilities of 0.91--0.95, and the readout baselines reached 0.92--0.96. Through Open Quantum, the baselines (0.92--0.98) and the two-qubit circuits from Qiskit (0.94--0.96) and PennyLane (0.91--0.96) also succeeded. Cirq's two-qubit circuits instead returned \texttt{00} for both marked states, with probabilities of 0.98 and 0.99. The failure was specific to the Cirq--Open Quantum pair: the same files ran correctly on the same device through the direct route.

The four-qubit circuits had low correct-answer probabilities on both routes (0.01--0.19, compared with 0.0625 for random guessing), so they did not provide a clear comparison. Removing $R_z$ gates predicts Cirq's wrong answers, but also predicts failures for PennyLane that we did not observe. The rule therefore does not explain all the results. Even without knowing how the service changed the circuits, the comparison located the fault in the Open Quantum route. The provider later identified its cause in its own code.

\section{Discussion and limitations}
\label{sec:discussion}

Although OpenQASM allowed us to exchange circuits between frameworks, we still had to account for differences in the gates they support and how they interpret them. In the Braket example (Section~\ref{sec:interop}), imported gate definitions introduced phase differences that affected controlled operations. So, even though we use QASM2 as the common format, it was not possible to guarantee that every framework executed the circuit in the same way.

For the hardware experiments, there was a limitation on circuit size and applications: we used small Grover circuits, adders, and individual gate checks. After the workaround for \texttt{Tuna-17}, the two-qubit Grover circuits worked well on all three devices. The four-qubit circuits, however, produced results close to random guessing on \texttt{garnet} and \texttt{Tuna-17}, making differences between builders difficult to identify. The defects we found show that the approach can help find bugs, but it is important to expand the evaluation to other applications and platforms, as well as to systems with more qubits.

As for the oracles, we used algorithms with known answers (adders and Grover), so we could check the answers directly, including cases where all implementations agreed. All three builders used Qiskit's transpiler, however, so the experiment did not compare different compilers. The direct route to \texttt{garnet} was also important: without it, we could not have distinguished a fault in Open Quantum from one in the device. These comparisons helped us find the three defects. To understand their causes, we examined the rotation gates and the adapter's gate declarations, and sent reproductions to Open Quantum, whose provider identified the parsing defect.

We ran several repeated batches under the same calibration. This was the case for three of the four repeats on \texttt{garnet} and three of the six on \texttt{ibm\_kingston}. The ten repeats therefore did not cover ten different calibration conditions. No calibration identifier was available for \texttt{Tuna-17}, so we could establish that its three repeats ran at different times, but not whether the calibration had changed.

An unexpected benefit of N$\times$M execution was that it helped us find defects in the backends. Targeted tests might have found the same defects, but initially we did not know which behaviors to test. The different implementations on different engines exposed disagreements and that triggered our investigation to further analyze the reason for it, with a specific combination builder-engine.

This investigation then led to the findings of three defects involving two providers and three different parts of the execution toolchain: execution consistent with reversed $R_x$ rotations on \texttt{Tuna-17}, an incorrect gate list in the Quantum Inspire Qiskit adapter, and a circuit-parsing defect in Open Quantum. The Open Quantum provider confirmed and fixed its defect. The two Quantum Inspire reports remain unconfirmed by the maintainers. These findings show the practical value of the comparisons and the variety of implementations, as they exposed behaviors that some circuits exercised and others did not, while the alternative execution routes helped us narrow down the source of the failures.

Finding these problems in a small set of experiments also raises questions about the reliability of quantum execution services, particularly when incorrect results arrive without an error message. Our experiments cannot establish how widespread such defects are, but they show that N$\times$M execution can help users assess these services and uncover problems that local simulation does not reveal. A further question is whether the same approach can expose defects in the interactions between processors, especially when a quantum solution implemented as a Quanifi flow combines components from different frameworks.

\subsection{Answers to the research questions}
\label{sec:rq_answers}

\begin{researchbox}
\textit{RQ1: what quantum programs can be expressed at the component level,
without writing framework-specific circuit code?}

\noindent\textbf{Answer.} We built the Grover flows by configuring and connecting processors, without writing circuit-level code. One table of inputs was used to build circuits with several frameworks and run them on several engines. The replication package~\cite{quanifiReplication} includes further NiFi flows for Deutsch--Jozsa, Bernstein--Vazirani, portfolio optimisation with QAOA, and maximum independent set, all executed using local Qrisp simulation. The independent-set example uses the graph and problem from the Classiq example~\cite{classiqMISExample}. These flows show which programs we could assemble from the available components. A user study is still needed to assess ease of use and whether this reduces development effort.
\end{researchbox}

\begin{researchbox}
\textit{RQ2: what does running every combination of framework implementation and
execution engine on real quantum computers reveal that comparing
implementations alone, or engines alone, does not, and what does it cost?}

\noindent\textbf{Answer.} N$\times$M execution helped us find three defects. Cirq's two-qubit circuits failed on \texttt{Tuna-17} but worked on the other devices, while the other builders worked on \texttt{Tuna-17}. Investigating this difference led us to the rotation defect and the adapter's incorrect gate declaration. For Open Quantum, circuits that failed through the service returned the correct answers when the same files were sent directly to \texttt{garnet}. Comparing both builders and execution routes gave us the evidence needed to distinguish these failures.

The additional executions are the main cost: each input requires $N \times M$ runs, plus any controls. In the three-device Grover experiment, each batch contained 20 circuits per device, including readout baselines and null replicates, with 1024 shots per circuit. We counted the circuit runs and shots, but did not measure their financial cost or the additional time required.
\end{researchbox}

\section{Conclusion and future work}
\label{sec:conclusion}

Quanifi lets users assemble algorithms from existing framework routines in one executable dataflow. N$\times$M execution helped us discover three defects in the execution toolchain: an incorrect gate-set declaration in the Quantum Inspire adapter, incorrect $R_x$ execution on Tuna-17, and circuit modification through Open Quantum. Comparing builders and execution routes exposed failures that we investigated with targeted tests. We reported all three defects and Open Quantum's provider confirmed and fixed its defect.

We compare Grover implementations by their probability of measuring the marked state. Other routines would need their own checks, such as the estimated phase for quantum phase estimation. Evaluating these checks and extending the matrix to hybrid workflows are future work.

Future work will extend the evaluation and the processor library. We plan a user study of Quanifi's accessibility, and we plan to support components written in other languages, such as C++ and Julia, so that Quanifi can reuse quantum routines beyond Python-based frameworks. The Q\#/QDK simulator processor developed in this work is a first step in this direction.

Quanifi also includes processors that mutate circuit code (OpenQASM) and processor configurations. In future work, we plan to combine these processors with probabilistic oracles and N$\times$M execution to study quantum software testing on real hardware. While this paper focused on composing and executing quantum programs, that study will evaluate how well the combined approach detects faults introduced into circuits and processor properties.

\section*{Acknowledgments}

This work was supported by the Ibero-American Programme for Science and
Technology for Development (CYTED), through RIPAISC -- the Ibero-American
Network for the Advancement of Quantum Software Engineering (525RT0174).

The authors also thank the support from the National Council for Scientific
and Technological Development (CNPq), Brasília, Brazil, grant number
406941/2025-4.

\bibliographystyle{IEEEtran}
\bibliography{references}

\end{document}